\documentclass[journal,transmag]{IEEEtran}
\usepackage{amsmath,graphicx,amssymb}
\usepackage{makecell}
\usepackage{multirow}
\usepackage{geometry}
\usepackage{verbatim}
\usepackage{xcolor}
\usepackage{url}
\usepackage{csquotes}
\usepackage{bm}
\usepackage{hyperref}
\title{Simplified magnet design and manufacture based on patterning of wide conductors}

\author{Diego Pereira Botelho\textsuperscript{1}, Victor Prost\textsuperscript{1},\\ Luana Barbosa Pina Pereira\textsuperscript{1}, Francesco A. Volpe\textsuperscript{1}\\
\textsuperscript{1}Renaissance Fusion, 38600 Fontaine, France \\
\thanks{Corresponding authors: Diego Pereira Botelho (email: \href{mailto:diego.pereira@renfusion.eu}{diego.pereira@renfusion.eu}) and Francesco A. Volpe (email: \href{mailto:francesco.volpe@renfusion.eu}{francesco.volpe@renfusion.eu})}}

\begin{document}

\maketitle

\begin{abstract}
The fabrication and assembly of High Temperature Superconducting (HTS) magnets can be significantly streamlined by (1) direct deposition of HTS onto modular components and (2) laser-ablated grooves to bound and guide the electric currents over the superconducting surfaces. \enquote{Coils} bounded by consecutive grooves can be individually powered, or connected in series, or in parallel. Applications include plasma-confinement devices such as stellarators and magnets for particle accelerators. 
On the way to HTS magnets generating more complicated three-dimensional fields, this paper validates the technique for two cylindrically symmetric magnets made of standard conductors. 
The optimized grooving pattern is considered as an inverse problem that is resolved using a least squares approach with Tikhonov regularization. This approach was first applied to design a magnet that replicates the magnetic field configuration of a microwave source of the gyrotron type. Gyrotrons require a particular profile of magnetic field, which our aluminium prototype successfully reproduced with 10$^{-1}$ precision. The second one, in copper, is a small-size, reduced-field magnet for Magnetic Resonance Imaging (MRI). This application requires highly uniform longitudinal fields. The achieved precision (about 50 ppm) is exceptional for an MRI magnet just getting out of the factory, and can easily meet hospital standards after minimal shimming or other error field correction.
Future work and other potential applications of wide, patterned conductors or superconductors are also discussed.
\end{abstract}

\section{Introduction}
Unlike the more widely adopted tokamak design, stellarators use a twisted three-dimensional magnetic field to confine fusion plasmas. The complex stellarator field provides advantages such as steady-state operation and improved stability, but introduces challenges in magnet design, manufacturing and assembling \cite{Gates2018}.

In first approximation, stellarator coils lie on a surface called Coil Winding Surface (CWS). 
Traditionally, stellarators adopt a {\em complex} three-dimensional CWS, reminiscent of a \enquote{conformal} surface corresponding to an inflated stellarator plasma boundary. On said 3D surface, the {\em simplest} filamentary coils that accurately generate the target field are sought \cite{Drevlak1998}. Such coils are relatively narrow and separated by {\em large gaps}~\cite{Grieger1992}. 

Renaissance Fusion \cite{renfusion2024} develops an innovative approach to simplify the fabrication of stellarator magnets. It envisions {\em simpler}, for instance piece-wise cylindrical CWSs, at the cost of more {\em complicated} current-patterns on them. 
Such cost, however, can be made negligible if one coats the CWS with HTS and then patterns it with an inexpensive system. Laser engraving of HTS, for instance, has been demonstrated using commercially available laser systems, achieving micro-metric precision. While earlier efforts primarily focused on reducing AC losses \cite{Nast2014,Prestigiacomo2017,Grilli2016, Pekarvcikova2023}, recent applications include specialized devices such as undulator magnets \cite{Krasch2024}. Other engraving, etching and machining techniques are possible as well. 
The result is a printed circuit where grooves guide the electric currents as to accurately generate the target field. Note that the current-carrying parts are wide, separated by {\em narrow gaps}, with the added benefits of better field utilization and reduced ripples.

In this paper, we take an intermediate step toward complex stellarator configurations by applying our wide conductor patterning technique to the design of axisymmetric magnets. Although simpler, these test cases effectively demonstrate the versatility and potential of this novel approach, including its applicability across a broad range of technologies. For the purposes of this proof of concept demonstration, standard resistive conductors (\textit{e.g.}, aluminium, copper) are used.

The paper is organized as follows. 
In Sec.~\ref{sec:Theory} we present the computational design tool and the underlying theory 
aimed at optimizing the laser engraving patterns to produce a given magnetic field. 
Secs.~\ref{sec:Gyrotron} and \ref{sec:MRI} describe the numerical modeling, prototype design and experimental results for two test cases: a magnet for a microwave source of the gyrotron type, 
and a magnet for magnetic resonance imaging (MRI). 
Future improvements and other potential axisymmetric applications are described in the Conclusions and Future Work section.

\section{Theoretical framework} 
\label{sec:Theory}
The theoretical basis underlying the proposed approach is introduced for axisymmetric field configurations. Therefore, CWSs will consist of cylinders with circular cross-sections.
 
\subsection{Problem formulation}
\begin{figure}
    \centering
    \includegraphics[width=1\linewidth]{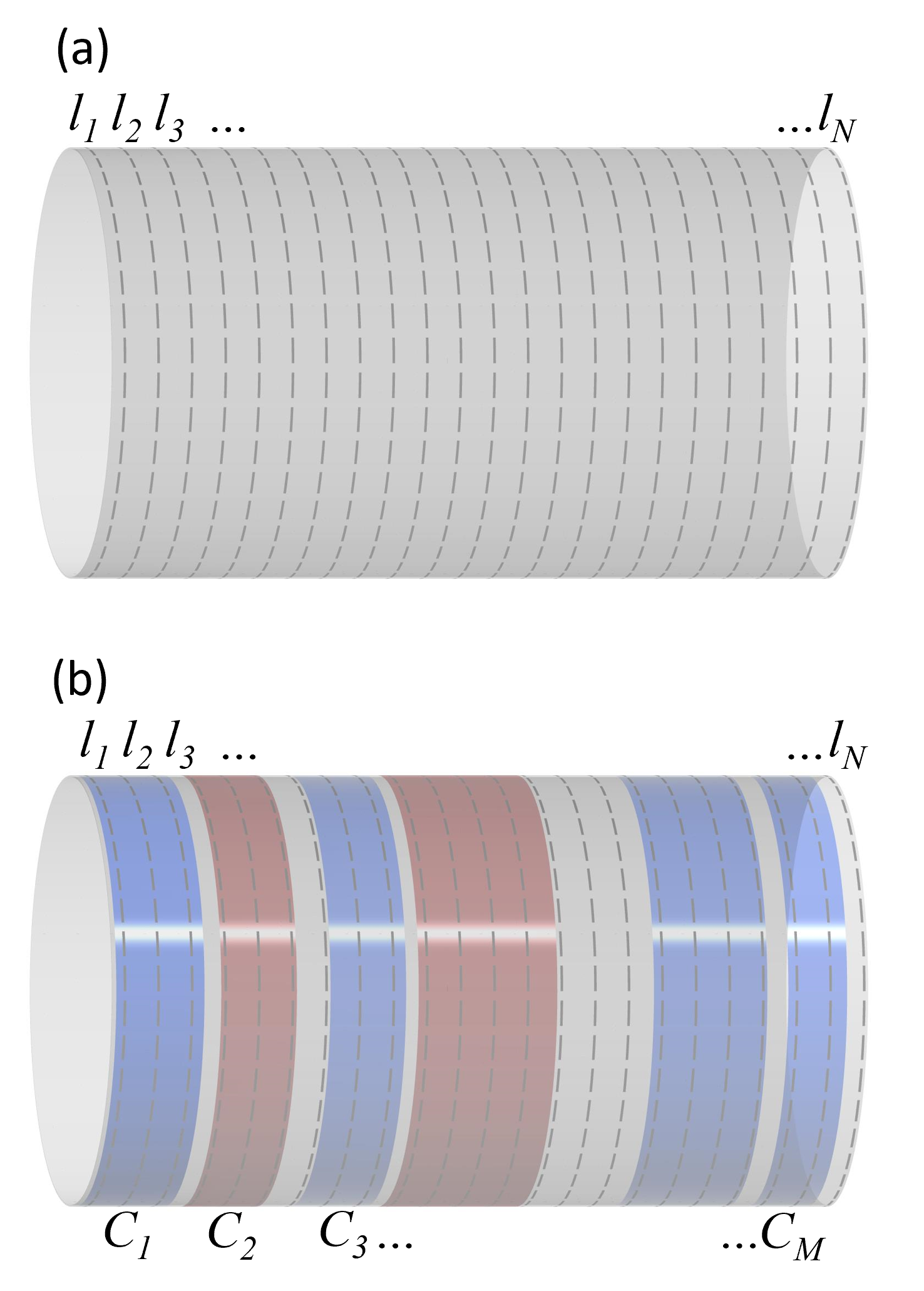}
    \caption{(a) Schematic of a cylindrical conductor as a set of $N$ current loops $l_1,\dots l_N$ of infinitesimal thickness carrying finite currents; (b) Grouping of the $N$ loops in $M < N$ coils $C_1, \dots C_M$ delimited by laser engravings (also represented as subsets of loops). The blue and red coloration denotes clockwise and counter-clockwise currents. Current-free gaps are shown in gray.}
    \label{fig:THEORY_loops & coils}
\end{figure}

In magnets producing axisymmetric field configurations, the current distribution over the CWS can only vary with respect to the position along the cylinder axis (there is no current variation with respect to the azimuthal angle). Thus, one can discretize the current over the CWS as a set of infinitesimal circular current loops (Fig.~\ref{fig:THEORY_loops & coils}a). The field produced by such current loops can be expressed analytically \cite{Simpson2001} and is used to calculate the magnetic flux density $\bm{B}$ at a given point $\bm{x}$ as the superposition of the field produced by all discretizing loops:
\begin{equation}
\bm{B}(\bm{x}) = \sum_{n=1}^{N} I_{ln} \bm{B}_{ln}(\bm{x}),
\end{equation}
where $I_{ln}$ is the current circulating through the loop $n$ and $\bm{B}_{ln}(\bm{x})$ the magnetic flux density produced by this loop \cite{Simpson2001}.

For a set of target points \(\bm{x}_p\), the equation above yields a linear system in which the loops’ currents are unknown. In matrix notation, we have:
\begin{equation}
\label{linear_system}
\bm{B}_P = \mathbf{B}_L \bm{I}_L,
\end{equation}
where $\bm{B}_P$ is the vector of magnetic flux density at the target points, \(\mathbf{I}_L\) is the vector of unknown currents $I_{ln}$ circulating through the corresponding loops and \(\mathbf{B}_L\) is a matrix defined as:
\begin{equation}
\mathbf{B}_L = \left(\bm{B}_{ln}(\bm{x}_p) \right)_{p,n},
\end{equation}
\noindent whose elements represent the magnetic flux density contribution from each discretizing current loop at each target point. This system allows for the determination of the current distribution required to generate the desired magnetic field at the specified target points.

\subsection{Electric current expansion in a polynomial basis}
In a cylinder aligned with the z-axis, one can express the current value $I_{ln}$ of the $n$-th current loop at the position $z_n$ as a polynomial of degree $d$:
\begin{equation}
\label{GYRO_poly expansion}
I_{ln} = \beta_0 + z_n \cdot \beta_1 + z_n^2 \cdot \beta_2 + \cdots + z_n^d \cdot \beta_d.
\end{equation}

In matrix notation, the polynomial expansion for the full set of current loops can be written as
\begin{equation}
\label{polynomial_expansion}
\bm{I}_L = \mathbf{Z} \bm{\beta},
\end{equation}
\noindent where $\mathbf{Z}$ is a Vandermonde matrix encoding the discretization of the magnet, being invertible if all $z_n$ values are distinct \cite{ciarlet1990handbook}, and $\bm{\beta}$ is the vector of polynomial coefficients.

Substituting Eq.~\eqref{polynomial_expansion} into Eq.~\eqref{linear_system}, we obtain:
\begin{equation}
\bm{B}_P = \mathbf{B}_L \mathbf{Z} \bm{\beta}.
\end{equation}
To solve directly for \(\bm{\beta}\), we can rely on a least squares formulation in which the objective is to minimize the difference between the target flux density \(\bm{B}_T\) at the set of points \(\bm{x_p}\) and the field produced by the model \(\mathbf{B_L} \mathbf{Z} \bm{\beta}\). The objective function is therefore:

\begin{equation}
\chi^2 = (\bm{B}_T - \mathbf{B}_L \mathbf{Z} \bm{\beta})^T (\bm{B}_T - \mathbf{B}_L \mathbf{Z} \bm{\beta}).
\end{equation}

\begin{comment}
To minimize it, we can take the derivative with respect to \(\bm{\beta}\) and set it to zero, leading to:

\[
\mathbf{Z}^T {\mathbf{B}_L}^T \mathbf{B}_L \mathbf{Z} \bm{\beta} = \mathbf{Z}^T {\mathbf{B}}_L^T \bm{B}_T.
\]

Assuming $\mathbf{Z}^T {\mathbf{B}_L}^T \mathbf{B}_L \mathbf{Z}$ is invertible, we can solve for \(\bm{\beta}\) as:
\end{comment}

To minimize it, we can take the derivative with respect to \(\bm{\beta}\) and set it to zero. We can then solve for \(\bm{\beta}\) as:

\begin{equation}
\bm{\beta} = (\mathbf{Z}^T {\mathbf{B}_L}^T \mathbf{B}_L \mathbf{Z})^{-1}\mathbf{Z}^T {\mathbf{B}}_L^T \bm{B}_T.
\end{equation}

This solution provides the coefficients \(\bm{\beta}\) which allow to reconstruct the optimized current distribution over the loops to approximate the target magnetic flux density \(\bm{B}_T\) over the set of points \(\bm{x_p}\). In this way, the problem of determining the coil currents that approximate a desired magnetic field can be treated as a polynomial regression problem, where the coefficients \(\bm{\beta}\) are found by solving the regularized least squares system.

% MOVED HERE TO APPEAR ON THE TOP OF PAGE 3
\begin{figure*}[ht]
\centering
\includegraphics[width=\linewidth]{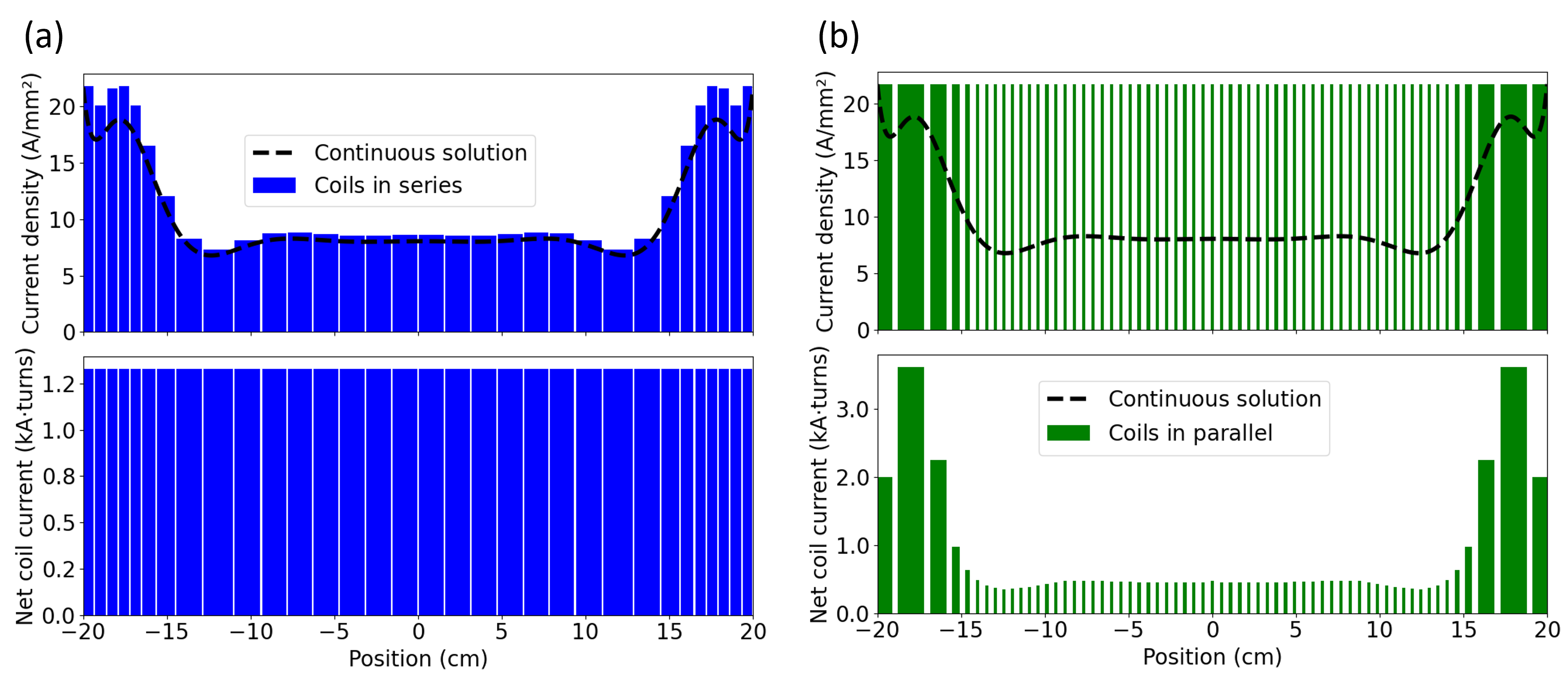}
\caption{Discretization of continuous current-density distributions (dashed black curves) into coil-currents separated by grooves. Each coil is modeled as a set of loops. (a) Series case: a single power supply (PS) feeds all coils, connected in series with each other, hence all coils carry the same current (bottom) but with different current densities (top). The area of a rectangle in the top corresponds to the coil current, and therefore all of them have the same area. The continuous current-distribution is approximated by strategically sizing and placing the coils (\textit{i.e.}, rectangles ranging from tall-and-narrow to flat-and-wide).
%This is an additive technique: coils are added at various locations. 
%each rectangle's area equals the current provided by the power supply; 
(b) Parallel case: a single PS feeds all coils, in parallel with each other, 
hence the current density is the same in all coils (top) but the coil currents are different (bottom). The target current distribution (dashed black line) is now obtained subtractively, by strategically sizing and placing current-free grooves that effectively, locally subtract current from the otherwise flat current density distribution.}
\label{fig:THEORY_discretization}
\end{figure*}

\subsection{Tikhonov regularization for stability}

To ensure stability, particularly when $\mathbf{Z}^T {\mathbf{B}_L}^T \mathbf{B}_L \mathbf{Z}$ is ill-conditioned, Tikhonov regularization can be applied to the least squares problem. The regularized version of the equation is:

\begin{equation}
\label{regularized_least_squares}
\bm{\beta} = (\mathbf{Z}^T {\mathbf{B}_L}^T \mathbf{B}_L \mathbf{Z}+ \lambda \mathbf{I})^{-1} \mathbf{Z}^T {\mathbf{B}_L}^T \bm{B}_T,
\end{equation}

\noindent where \(\lambda\) is the regularization parameter and \(\mathbf{I}\) is the identity matrix. In practice, the regularization term also indirectly controls the maximum current density required to approximate the target field for a given precision.

The effect of the conductor volume can be easily integrated into the system by the introduction of current loops of different radii spanning the radial thickness of the conductor ( \textit{i.e.}, a radial discretization additional to the axial one presented here). Loops with different radii but with the same $z_n$ can be lumped into the same element of the matrix \(\mathbf{B}_L\).

\subsection{Coils in series \textnormal{vs} coils in parallel}

To the framework presented above, one can introduce the concept of coils (\textit{i.e.} conductive bands bounded by grooves). The relation between coils and current loops is illustrated in Fig.~\ref{fig:THEORY_loops & coils}b. Accounting for these coils requires the introduction of an incidence matrix $\mathbf{W}_C$ to Eq.~\eqref{linear_system}. The incidence matrix $\mathbf{W}_C$ is sparse and its non-zero elements along each row are adjacent and identical to each other, being defined as $1$ divided by the number of loops comprised by a given coil:

\[
\mathbf{W}_C =
\begin{bmatrix}
\frac{1}{|\bm{C}_1|} \mathbf{1}_{|\bm{C}_1|} & 0 & \cdots{}& 0 \\
0 & \frac{1}{|\bm{C}_2|} \mathbf{1}_{|\bm{C}_2|} & \ddots{}& \vdots{} \\
\vdots & \ddots & \ddots& 0 \\
0 & \cdots{}& 0& \frac{1}{|\bm{C}_M|} \mathbf{1}_{|\bm{C}_M|}
\end{bmatrix},
\\[20pt]
\]

\noindent where:
\begin{itemize}
    \item \( \mathbf{W}_C \) is an \( M \times N \) matrix, with \( M \) representing the number of coils and \( N \) representing the number of current loops in the system;
    \item \( |\bm{C}_m| \) is the number of current loops that comprise the \( m \)-th coil, \( C_m \);
    \item \( \mathbf{1}_{|\bm{C}_m|} \) is a row vector of ones of length \( |\bm{C}_m| \), corresponding to the current loops within the \( m \)-th coil.
\end{itemize}

For a known distribution of coils $\mathbf{W}_C$, the independent coils’ currents that approximate the target field can be obtained by the resolution of the linear system:
\begin{equation}
\bm{B}_T = \mathbf{B}_L \mathbf{W}_C \bm{I}_C,
\end{equation}
\noindent where $\bm{I}_C$ is a vector of coil currents.

Alternatively, for a given current distribution obtained with Eq.~\eqref{regularized_least_squares}, one can compute a distribution of grooves — and therefore coils — that approximates the continuous current solution. This calculation can be made following two simplifying situations: coils in series or coils in parallel.

In a series arrangement, since the net current provided by the power supply (PS) is the same for all individual coils, their current densities can be varied according to their respective widths (\textit{i.e.}, distance between neighboring grooves). Essentially, this is a discretization procedure which is somewhat similar to a trapezoidal integration method. The key difference is that each \enquote{trapezoid} is designed to have an equal area, corresponding to the coils’ current (Fig.~\ref{fig:THEORY_discretization}a).

In the case of a parallel arrangement, coils are subjected to a single potential difference provided by the PS. Consequently, this arrangement generates a single current density for all coils. Thus, one can consider that the widths an the number of grooves become the main degrees of freedom to modulate the local effect of the current density distribution (Fig.~\ref{fig:THEORY_discretization}b).

Even though these two arrangements have been proven to be numerically equivalent, sensitivity analyses have shown that parallel arrangements are oversensitive to mechanical inaccuracies and/or thermal effects impacting on the resistance of each coil. Indeed, parallel arrangements rely on a very precise control of the copper foil’s electric, physical and thermal properties to work as expected. For this reason, only coils arranged in series were considered for the experimental devices described below.

The discretization processes for series and parallel arrangements can be treated iteratively, by sweeping the combinations of feasible coils and grooves widths and distributing them in such a way as to approximate both the local value of the underlying continuous current distribution and its overall integral over the full magnet. In the test cases addressed in this work, the degradation of the field precision due to discretization was of approximately one order of magnitude. Examples of discretization of a given continuous current distribution solution in terms of series and parallel arrangements are illustrated in Fig.~\ref{fig:THEORY_discretization}.

\section{Gyrotron magnet}
\label{sec:Gyrotron}

\begin{figure}[ht]
\centering
\includegraphics[width=\columnwidth]{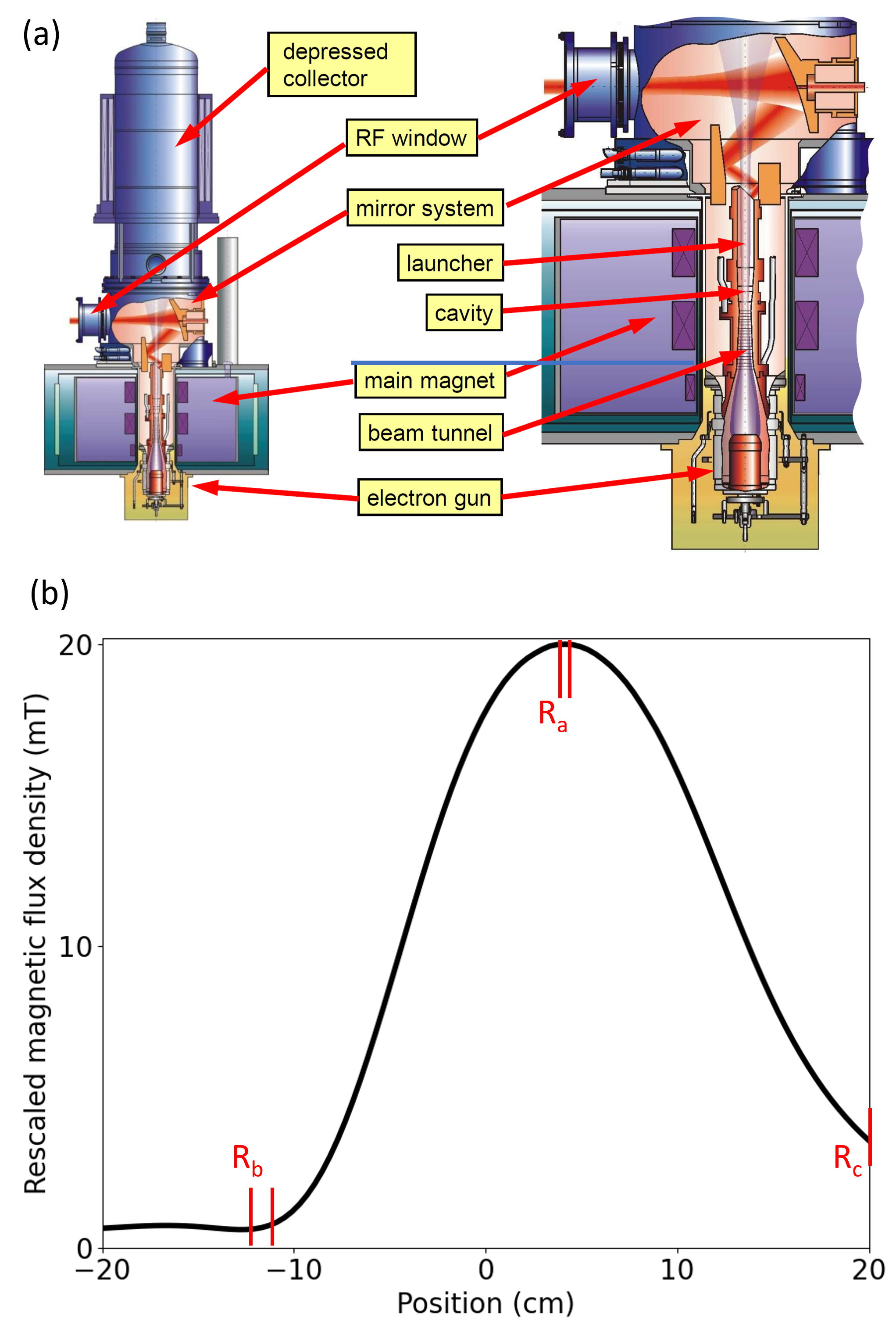}
\caption{(a) Schematic of a typical gyrotron heating source for fusion plasmas (from \cite{Jelonnek2017}). (b) Scaled gyrotron magnetic flux density magnitude profile along the magnet axis. The three regions \textbf{R$_{\text{a}}$}, \textbf{R$_{\text{b}}$} and \textbf{R$_{\text{c}}$} correspond respectively to the cavity, the electron gun and the launcher in Fig.~a.}
\label{fig:GYRO_device scheme & target field profile}
\end{figure}

Two test cases consisting of reduced-field versions of magnets used in commercial devices are considered in this study. The first magnet is designed to replicate a typical gyrotron field profile \cite{Jelonnek2017}. In gyrotrons, a well-defined magnetic field profile along the bore is necessary to guide the electron beam along the device’s axis and determine the cyclotron frequency, allowing frequency tuning and output power control.
Figure~\ref{fig:GYRO_device scheme & target field profile} shows a schematic of the gyrotron device considered as reference here and the required magnetic field profile along the magnet's axis \cite{Jelonnek2017}. Table~\ref{GYRO_field_profile_requirements} lists the gyrotron magnetic field density requirements adapted to the context of our prototype magnet which is 40 cm long with a 20 cm internal diameter, and scaled-down peak magnetic flux density of 20 mT. The regions $\mathrm{R_a}$, $\mathrm{R_b}$ and $\mathrm{R_c}$ are indicated in Fig.~\ref{fig:GYRO_device scheme & target field profile}b.

\begin{table}[h!]
\centering
\caption{Target field ranges (\cite{Jelonnek2017} scaled to a peak of 20 mT).}
\label{GYRO_field_profile_requirements}
\begin{tabular}{ccc}
\hline
\textbf{Region} & \textbf{Min (mT)} & \textbf{Max (mT)} \\
\hline
$\mathrm{R_a}$ & 19.98 & 20.02 \\
$\mathrm{R_b}$ & 0.4 & 1.0 \\
$\mathrm{R_c}$ & 2.4 & - \\
\hline
\end{tabular}
\end{table}

%Next, optimization results for the gyrotron field profile are presented.

\subsection{Numerical results}
\begin{figure}[ht]
\centering
\includegraphics[width=\columnwidth]{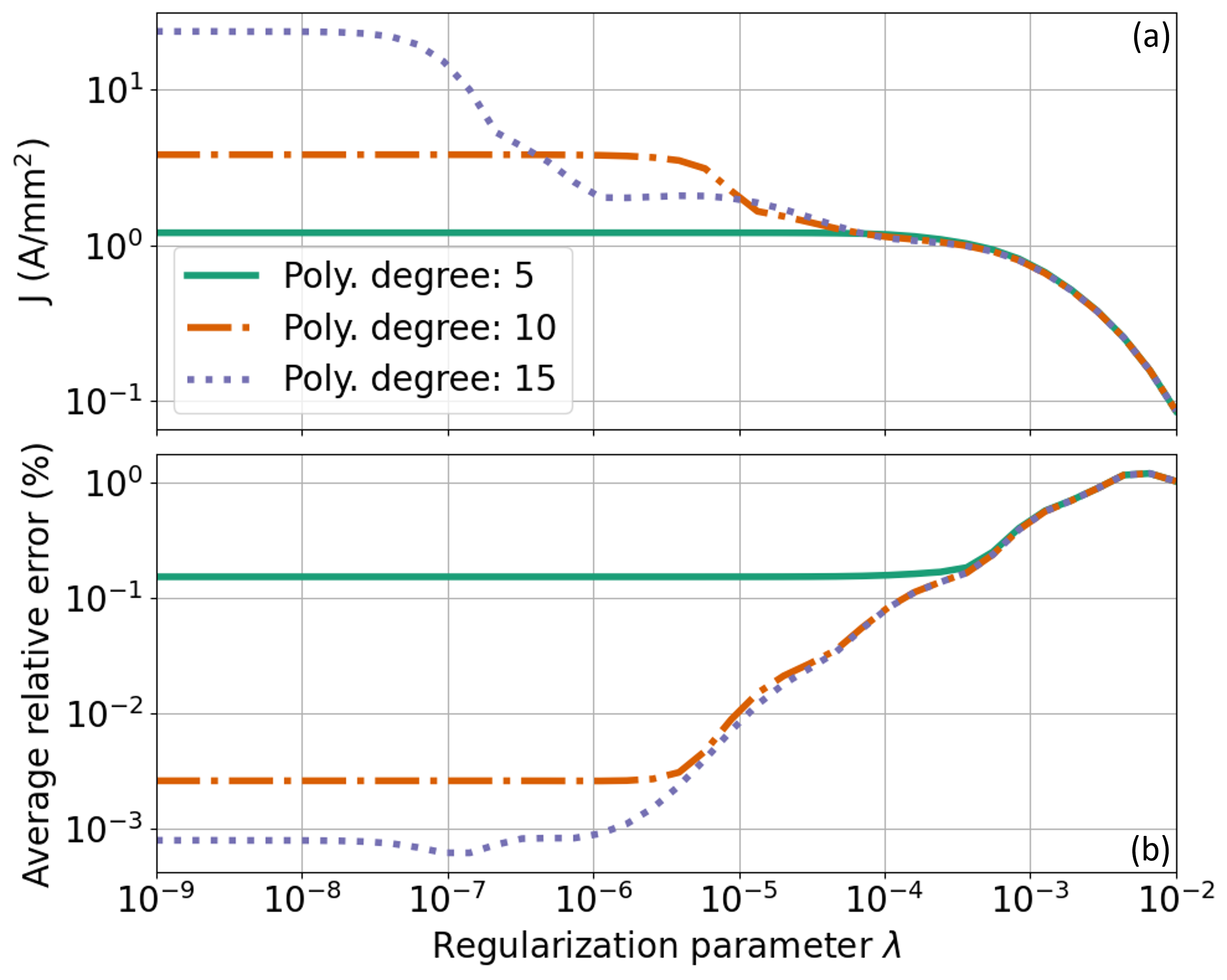}
\caption{(a) Maximum current density and (b) average relative local error as functions of the Tikhonov regularization parameter $\lambda$ for three polynomial degrees $d$: 5, 10 and 15 (Eq.~\eqref{GYRO_poly expansion}).}
\label{fig:GYRO_convergence wrt regularization}
\end{figure}

Optimization results for the gyrotron field configuration are shown in Fig.~\ref{fig:GYRO_convergence wrt regularization}, which highlights the convergence of the continuous problem with respect to the regularization parameter and polynomial degree. It is interesting to note that, for a given polynomial degree, the regularization indirectly controls the maximum current density required to approximate the target field as well as the precision of the approximation. The average relative error $\epsilon_{\text{avg}}$ in the plot is defined as follows:

\[
\epsilon_{\text{avg}} = \frac{1}{P} \sum_{p=1}^{P} \frac{\| \bm{B}_t(\bm{x}_p) - \bm{B}_m(\bm{x}_p) \|}{\| \bm{B}_t(\bm{x}_p) \|},
\]
\noindent where ${P}$ is the number of target points $\bm{x}_p$, $\bm{B}_t$ and $\bm{B}_m$ are target and magnet (approximated) flux densities, respectively, and $\| \cdot \|$ is the ${l^2}$ norm. One can additionally introduce the maximum local error $\epsilon_{\text{max}}$ as:
\[
\epsilon_{\text{max}} = \max_{p = 1, \dots, P} \left( \frac{ \| \bm{B}_t(\bm{x}_p) - \bm{B}_m(\bm{x}_p) \| }{ \| \bm{B}_t(\bm{x}_p) \| } \right).
\]

\begin{figure}[ht]
\centering
\includegraphics[width=\columnwidth]{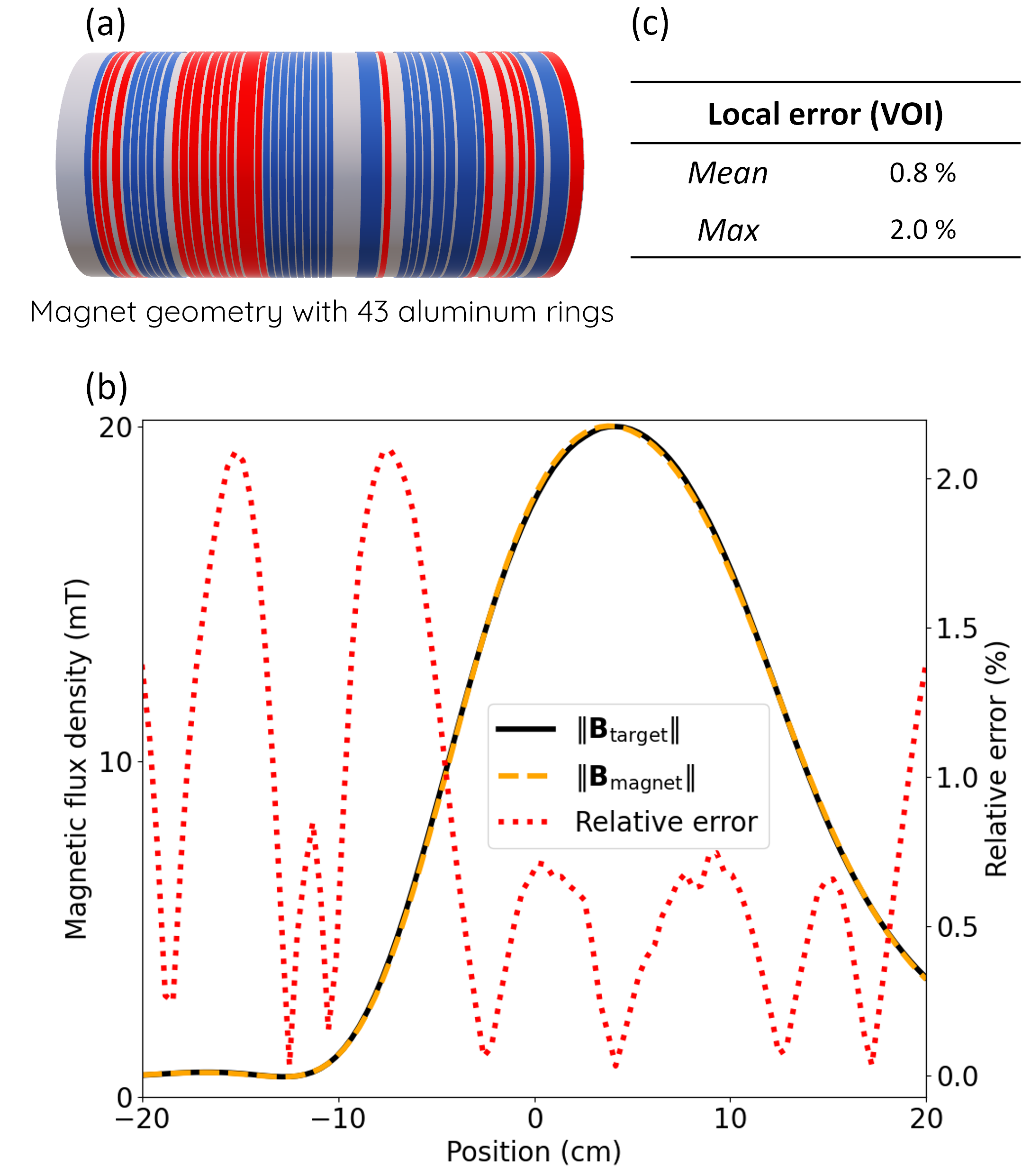}
\caption{(a) Grooved aluminium conductor optimized to replicate the gyrotron magnetic field profile; the design consists of 43 aluminium coils in series with each other, with the blue and red colors 
representing positive and negative currents. (b) Target magnetic flux density profile (black) and designed magnet computed profile (orange dashed, vertical axis on the left). The relative error is shown as a red dotted line. (c) Local errors in the volume of interest (VOI), which closely covers the magnet axis.}
\label{fig:GYRO_gyrotron_design}
\end{figure}

Figure~\ref{fig:GYRO_gyrotron_design} shows the optimized gyrotron magnet design as well as the comparison between the approximated and target magnetic field profiles along the axis. Geometric constraints were imposed on the design of this magnet to simplify manufacturing of the prototype parts as well as their assembly. Notably, coils and grooves’ widths were constrained to be multiples of 1 mm. Nevertheless, one can observe an agreement between the magnet and target flux densities with a maximum local error of 2 \% and an average error of 0.8 \%.

\subsection{Error sensitivity analysis}
A sensitivity analysis was carried out to investigate the impact of systematic coil and groove width inaccuracies. Indeed, since the prototyping approach in this case relies upon the principle of stacked aluminium and plastic plates (Sec.~\ref{sec:GYRO_experimental validation section}), systematic errors in plates’ widths tend to have a cumulative effect along the magnet. Furthermore, mechanical tolerances on the aluminium and plastic plates were specified during manufacturing within 0.1 mm. Table~\ref{tab:GYRO_systematic innacuracy error sensitivity} shows the impact of these mechanical tolerances on field precision.

\begin{comment}
\begin{figure}[ht]
\centering
\includegraphics[width=\columnwidth]{figures/GYRO_error_analysis.PNG}
\caption{(a) Impact of random coil width inaccuracies on the precision of the magnet field. (b) Comparison between target and approximated fields along the axis for the five cases (C1 to C5) described in Table \ref{tab:error}.}
\label{fig:GYRO_error_analysis}
\end{figure}
\end{comment}
~
\begin{table}[ht]
\centering
\caption{Impact of systematic aluminium and plastic width inaccuracies on magnetic field precision.}
\label{tab:GYRO_systematic innacuracy error sensitivity}
\begin{tabular}{cccc}
\hline
\textbf{Al var. (mm)} & \textbf{Plastic var. (mm)} & \textbf{Max error} & \textbf{Avg error} \\
\hline
0   & 0   & 2.0 \%  & 0.8 \% \\
0.1 & 0   & 19 \% & 3.4 \% \\
0   & 0.1 & 27 \% & 4.5 \% \\
0.1 & 0.1 & 48 \% & 8.0 \% \\
\hline
\end{tabular}
\end{table}

Due to the cumulative effect of systematic errors, even an increase of 0.1 mm in the width of aluminium plates can lead to local errors in the order of 19 \%. It is interesting to highlight however that local relative errors tend to be much higher on the tail of the field profile, where the field magnitude is much lower than elsewhere.

\subsection{Experimental validation}
\label{sec:GYRO_experimental validation section}
\begin{figure}[ht]
\centering
\includegraphics[width=\columnwidth]{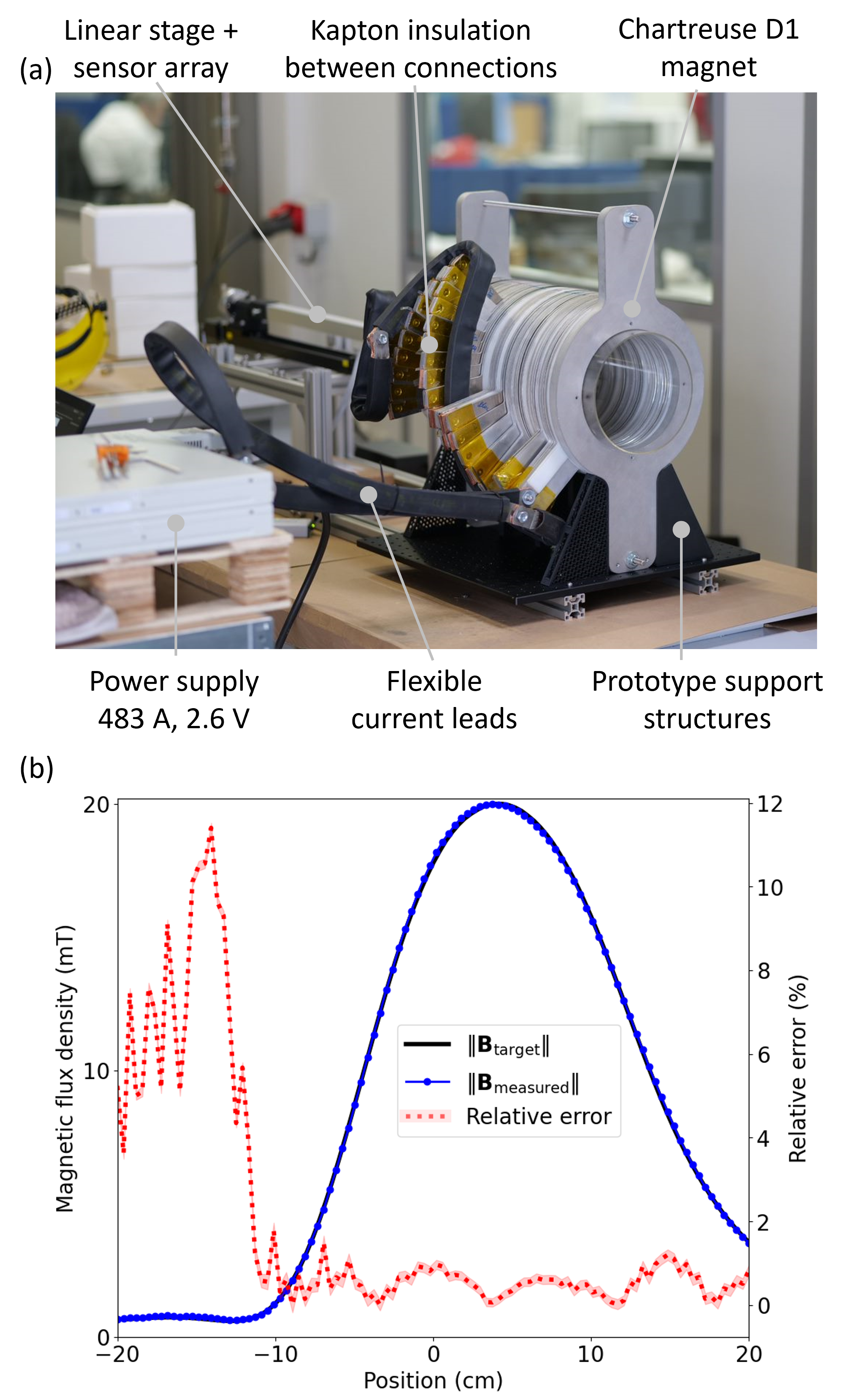}
\caption{(a) Gyrotron magnet and test bench. (b) Comparison between the target (black line) and measured (blue line and markers) magnetic fields along an axial trajectory, together with the relative error between the two (red dotted line surrounded by a shaded area representing measurement uncertainty). The discrepancy between the 
measurements and the target here is comparable with the discrepancy between 
prediction and target (Fig.~\ref{fig:GYRO_gyrotron_design}b) in the cavity region around the field peak, but is higher at low fields, near the electron gun.}
\label{fig:GYRO_prototype}
\end{figure}

To demonstrate the ability of the proposed magnet design approach to accurately replicate predetermined magnetic field configurations, a proof of concept prototype of the gyrotron test case was constructed (Fig.~\ref{fig:GYRO_prototype}a). Since the automated system allowing the laser-engraving and the winding of a (super-)conductor foil over a cylindrical substrate was still under development at the time of this experiment, aluminium annular discs juxtaposed with plastics annular discs of varying widths were used to construct the magnet and  represent the conductors bounded by straight corrugations as shown in Fig.~\ref{fig:GYRO_gyrotron_design}. The coil geometric sizes, their radial thicknesses and minimum width were dimensioned to use a PS of 10 V and 1 kA, while limiting the maximum temperature increase in the magnet to 50~°C over an experimental measurement time of 5~min. The aluminum discs included two straight radially outwards segments (current leads) used to connect the discs in series with each other and allow for changing the current direction (red-blue coils shown in Fig.~\ref{fig:GYRO_gyrotron_design}a). Lastly, flexible braided current leads (IBSBADV240, Amperio, Switzerland) were used to connect the PS to the prototype while closely following the trajectory of the aluminium leads. This enabled  reducing the impacts of the spurious fields produced by the interconnections of aluminum leads. 

An experimental stage composed of a hall-effect magnetic sensor array (Hallinsight, MetroLab, Switzerland) mounted on a linear and rotary stage was used to map the magnetic field within the magnet. Measured data was post-processed to remove the background field, and align the z-axis between the magnet and the experimental stage. Figure~\ref{fig:GYRO_prototype}b shows the comparison between measured and target field magnitudes along an axial line 15 mm away from the center of the magnet axis. The target and measured profile are in agreement with an average error of 2~\% and a maximum local error of about 11~\% on the tail of the distribution. Sensitivity analyses have shown that the higher error at the low field area is due to both the low field values in this region of the profile, the inherent mechanical inaccuracies of this particular prototyping method (stacked discs), and the influence of the field produced by the current leads.

\section{MRI magnet}
\label{sec:MRI}
The second test case consists of a magnet designed to provide a highly homogeneous field for MRI systems. Field homogeneity in the order of few parts-per-million is a prerequisite for reliable MRI signal detection \cite{Manso2023}. MRI systems utilize a combination of magnets to excite and detect the resonance of hydrogen nuclei within the human body, enabling the synthesis of detailed 3D images. The primary component of an MRI system is a magnet that generates a homogeneous and often strong magnetic field within a volume of interest (VOI). The body part to be imaged is positioned within the VOI, typically defined in terms of diameter of spherical volume (DSV). Field homogeneity is one of the key parameters of MRI systems, as it is directly related to imaging quality. In general, it is quantified in terms of its volumetric root-mean-square value ($V_{\text{RMS}}$) in parts-per-million (ppm) according to the following relation:

\[
V_{\text{RMS}} = \frac{1}{V} \int_V \left[ \Vert \bm{B}(\bm{x}) \Vert - \Vert \bm{B}(\bm{x}_0) \Vert \right]^2 dV,
\]

\noindent where \(V\) is the volume of interest and $\bm{B}(\bm{x})$ and $\bm{B}(\bm{x}_0)$ are the magnetic field densities at a point $\bm{x}$ and at the center of the VOI, respectively.

MRI magnets vary depending on the manufacturer and the specific requirements of the application. A compilation of parameters for the main MRI magnet including dimensions, field strengths, and field homogeneity can be found in \cite{Manso2023}. According to this compilation, field homogeneity in commercial MRI systems is typically smaller than 5~ppm for a 40~cm DSV. In terms of dimensions, it is important to note that the bore size indicated in \cite{Manso2023} corresponds to what is commonly referred as warm bore, \textit{i.e.}, the free space available within the device for the patient. Based on information provided by MRI manufacturers, typical dimensions for the main MRI magnet are in the range of 1 meter for the \enquote{cold} bore diameter and 2 meters for the magnet length.

An MRI magnet based on laser patterned conductors was developed through the combination of the coil optimization algorithm described in Sec.~\ref{sec:Theory}, electromagnetic finite element modelling and mechanical design. Optimization parameters were fine-tuned with the finite element model analysis, as well as the mechanical design and manufacturing constraints. Similar to the gyrotron magnet prototype, the MRI magnet was designed with a length of 40 cm and an internal diameter of 20 cm, sufficient for medical imaging of limbs or small animals. Compared to typical MRI, however, magnetic flux density was significantly reduced to 0.1~T.

The main specifications of the prototype are shown in Table~\ref{MRI_specifications}. The prototype dimensions correspond to a typical MRI magnet scaled down by a factor~5.

\begin{table}[h!]
\centering
\caption{Scaled-down MRI magnet specifications}
\label{MRI_specifications}
\begin{tabular}{lc}
\hline
\textbf{Dimensions} & Ø 20 cm, L 40 cm \\
 \textbf{Volume of Interest (DSV)}&Ø 8 cm\\
\textbf{Field} & 100 mT peak field \\
\textbf{Corrugations} & Laser engraved \\
\textbf{Target} & $\varepsilon_{\text{max}} \approx 10^{-5}$ \\
\textbf{Meas. tech.} & Linear stage and NMR probe \\
\textbf{Material} & Cu – PET sheet \\
\textbf{Temperature} & 77 K \\
\textbf{Pressure} & atm \\
\hline
\end{tabular}
\end{table}

\subsection{Numerical results}
\begin{figure}[ht]
\centering
\includegraphics[width=\columnwidth]{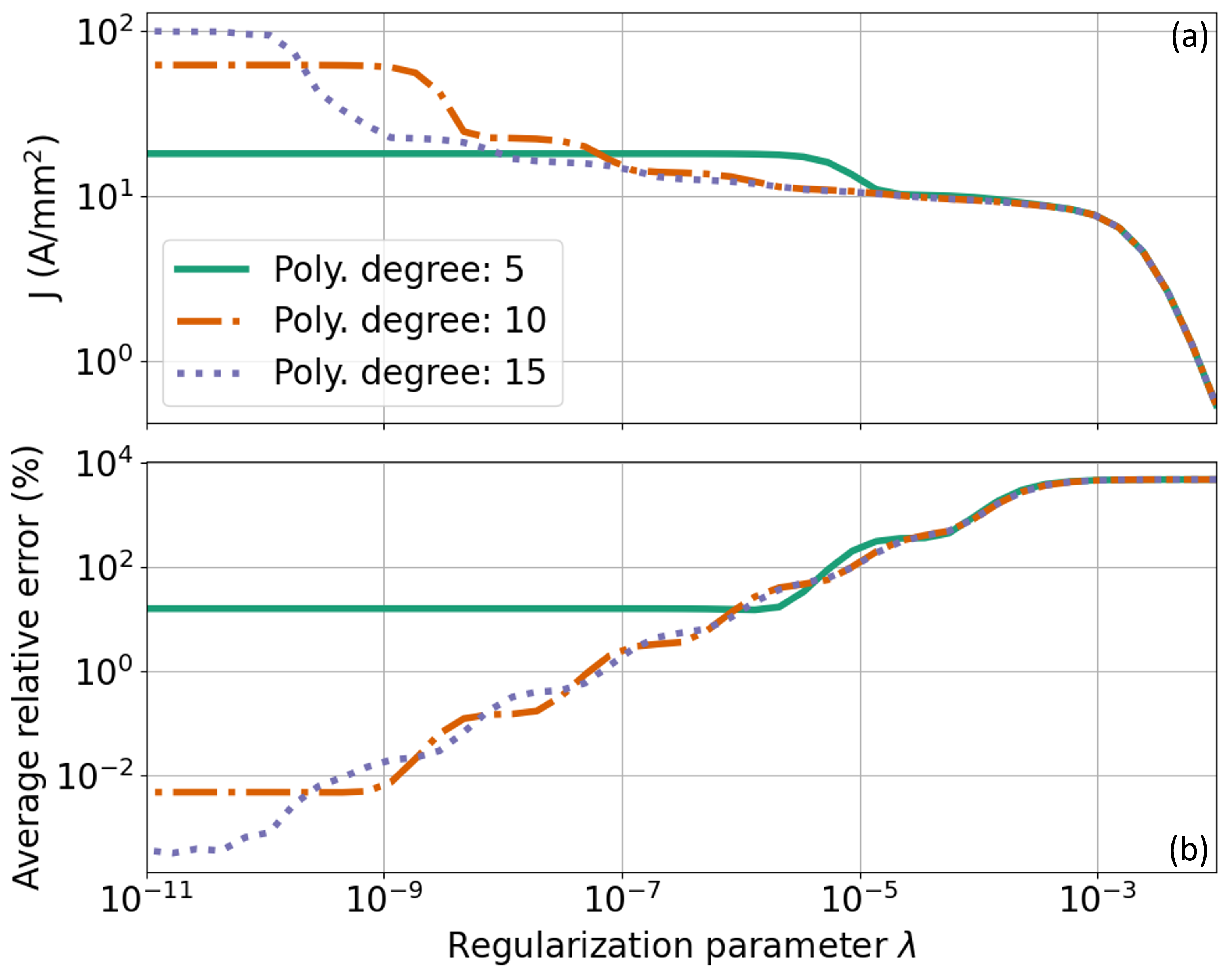}
\caption{Maximum current density magnitude (a) and  \(V_{RMS}\) homogeneity (b) as functions of the Tikhonov regularization parameter $\lambda$ for three polynomial degrees $d$: 5, 10 and 15 (Eq.~\eqref{GYRO_poly expansion}).}
\label{fig:MRI_numerical convergence}
\end{figure}

The engraving configuration was optimized to ensure the required MRI field uniformity. Figure~\ref{fig:MRI_numerical convergence} shows the convergence of the continuous problem with respect to the regularization weight.

\begin{figure}[ht]
\centering
\includegraphics[width=\columnwidth]{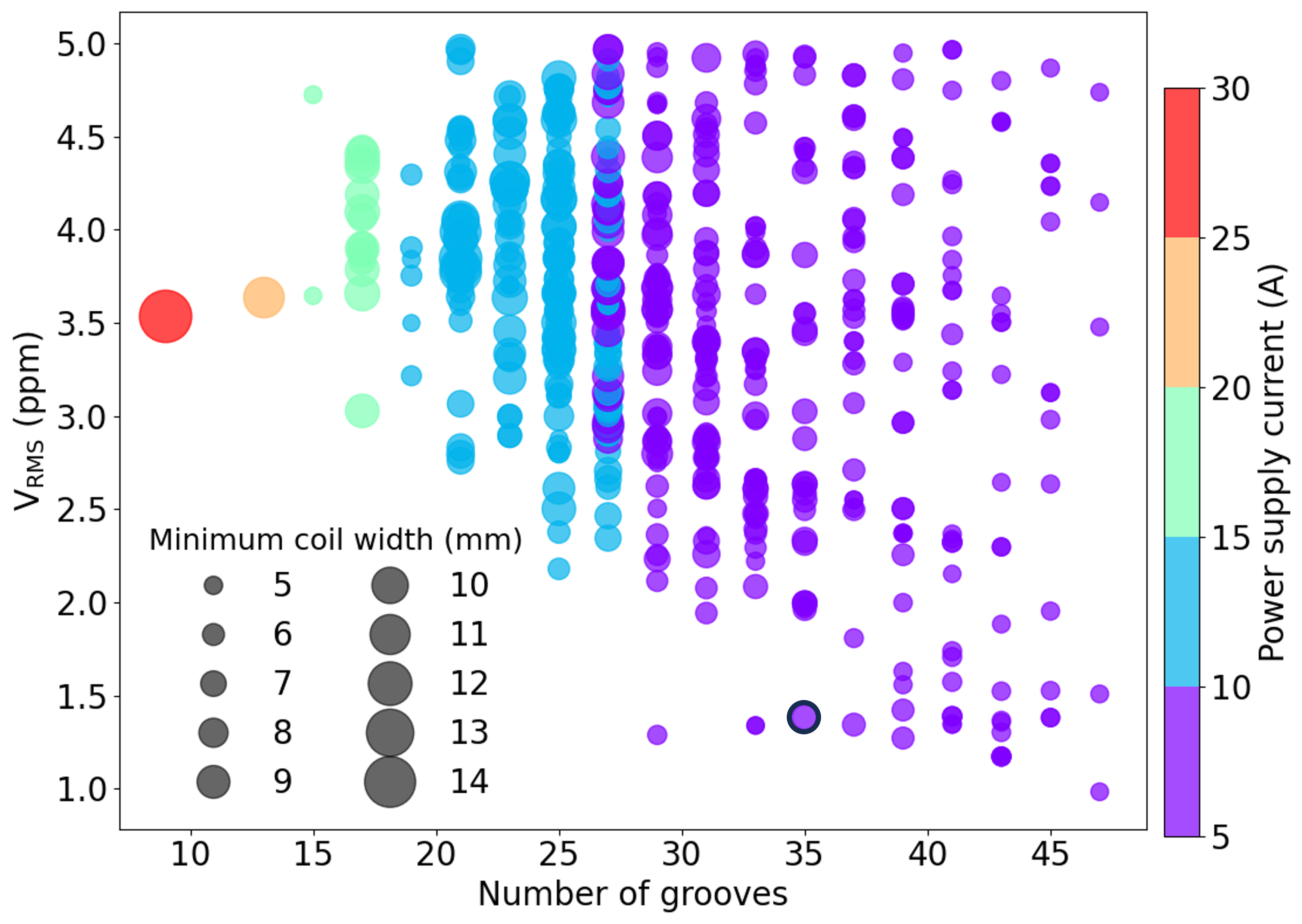}
\caption{
Optimized designs database for MRI magnets achieving the target $V_\text{RMS}$ for various numbers of grooves, minimal coil-widths, and power supply currents. 
%Solution set for a range of number of grooves, minimum coil widths and power supply current considering (150 turns). 
The selected design ($V_\text{RMS} = 1.38$ ppm, 35 grooves) is highlighted with a black circle.
}
\label{fig:MRI_design_constellation}
\end{figure}

The developed coil design framework can be used to generate a set of design solutions spanning a large range of constructive parameters, providing a great flexibility with respect to prototyping constraints. Figure~\ref{fig:MRI_design_constellation} shows an example of a set of design solutions (homogeneity \(V_{RMS}\) $\leq$  5 ppm) for a prototype of 150 windings with varying number of grooves, minimum coil widths and applied currents. The choice of this number of windings was made as it offered a good balance between thermal effects and material usage.

\begin{figure}[ht]
\centering
\includegraphics[width=\columnwidth]{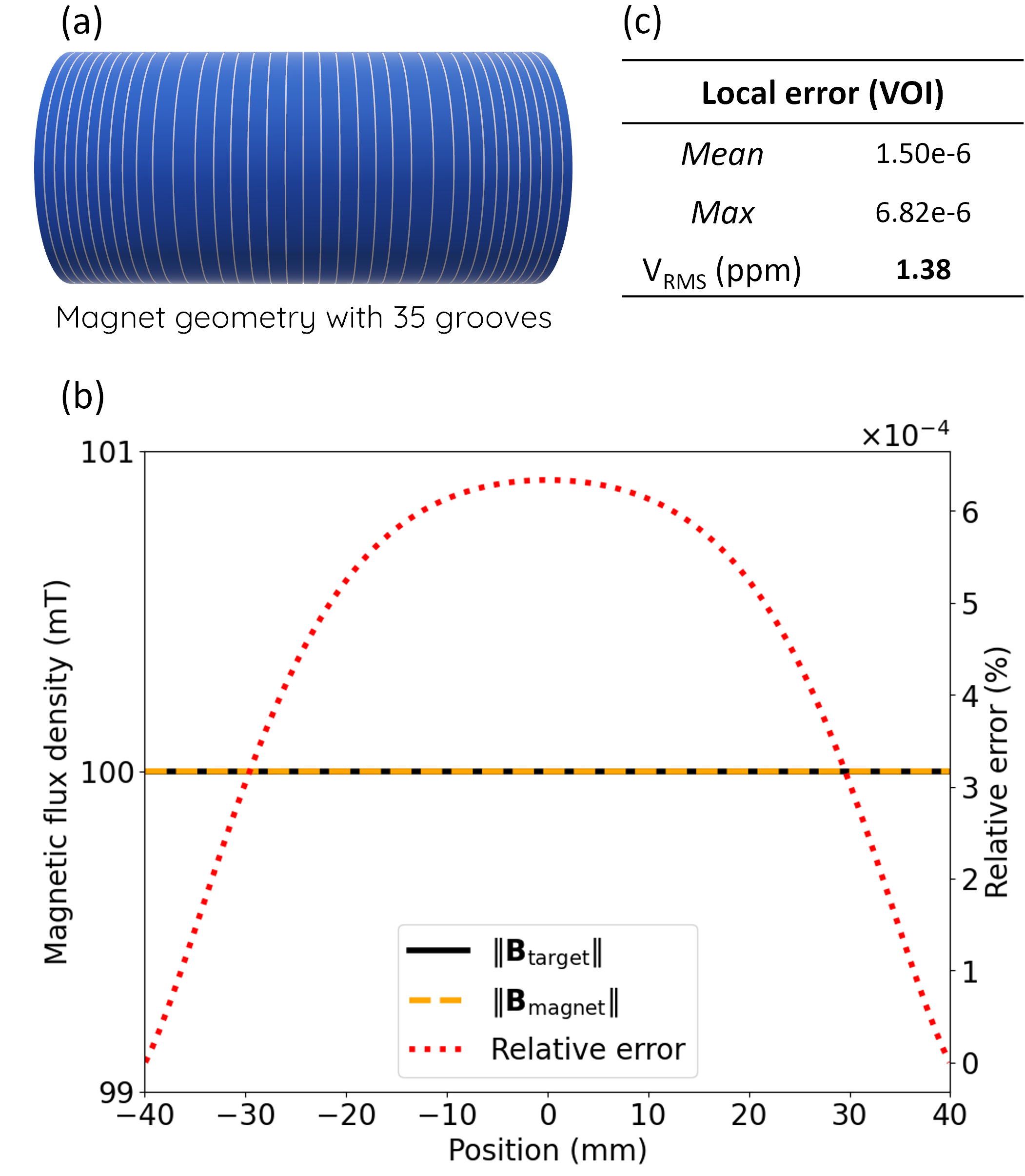}
\caption{Straight grooved copper conductor optimized to replicate the MRI magnetic field configuration. (a) Magnet configuration: 36 coils with widths spanning from 6.8 mm to 15.3 mm separated by 1 mm grooves. (b) Calculated (dashed orange) and target (solid black) magnetic flux density profile along the axis with the resulting error calculation (dotted red line). (c) Characterization of the homogeneity within the VOI.}
\label{fig:MRI_design}
\end{figure}

Figure~\ref{fig:MRI_design} shows the chosen magnet configuration as well as the characterization of the field homogeneity within the volume of interest (VOI). This configuration consists of 36 conductive stripes (or coils) with widths spanning from 6.8 mm to 15.3 mm separated by 1 mm grooves. In this design, the different coils are arranged in a series configuration. One can notice that a field homogeneity \(V_{RMS}\) of 1.38 ppm was computed (Fig.~\ref{fig:MRI_design}c), which well within the commercial MRI range (\(V_{RMS}\) $\leq$ 5 ppm). Fig.~\ref{fig:MRI_design}b shows in more details the comparison between the target and the theoretical magnetic flux density along the axis within the volume of interest, in which one can observe that the maximum relative error is in the order of 6 x 10\textsuperscript{-6}.

\begin{figure}[ht]
\centering
\includegraphics[width=\columnwidth]{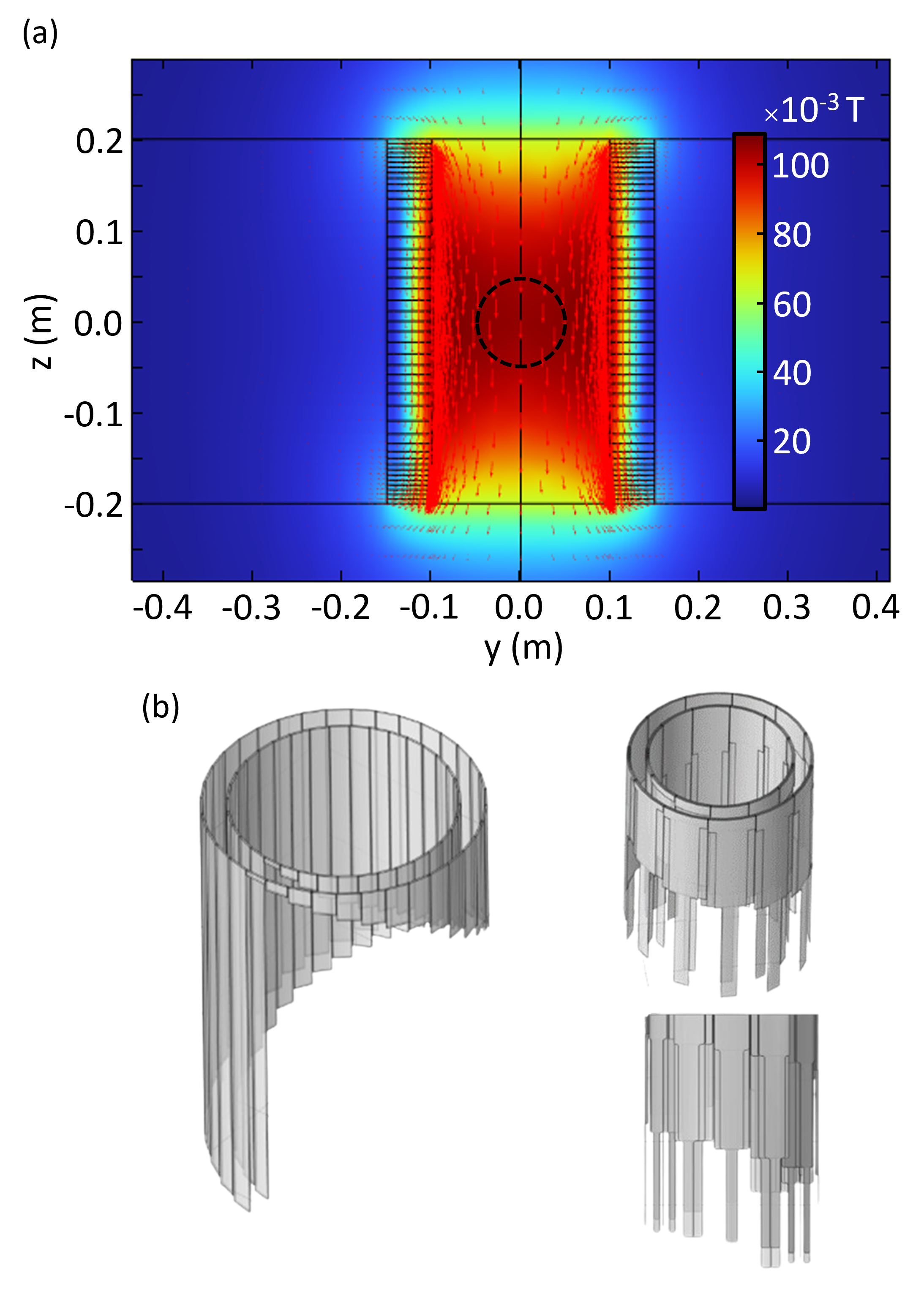}
\caption{(a) COMSOL® simulation of the MRI magnet with the shaded circle showing the volume of interest in which field homogeneity was calculated and the color plot representing the magnetic flux density magnitude. (b) Example of \enquote{staircase} current leads designs investigated to minimize their impacts on field homogeneity.}
\label{fig:MRI_fem_validation&leads}
\end{figure}

Finite Element Method (FEM) simulations in COMSOL® were performed to validate the developed numerical tools and investigate various current leads design and the impact of thermo-mechanical effects in the prototype. An initial 3D MRI model was constructed in COMSOL® with a given coil radial thickness but without any leads in order to validate the analytical models used for the groove optimization. The results from the COMSOL® simulation, shown in Fig.~\ref{fig:MRI_fem_validation&leads}a, demonstrated a field uniformity of \(V_{RMS}\)~=~1.23~ppm, validating the optimized groove pattern generated by our magnet design tools.

After validating the ideal magnet configuration (Fig.~\ref{fig:MRI_fem_validation&leads}a), various current leads geometries were included in the model (Fig.~\ref{fig:MRI_fem_validation&leads}b) to investigate their impact on the field homogeneity within the volume of interest. Initial current leads configurations (left graphic in Fig.~\ref{fig:MRI_fem_validation&leads}b) perturbed the field quality by \(V_{RMS}\) $\simeq$ 30~ppm. Optimizing leads geometries (bottom right graphic in Fig.~\ref{fig:MRI_fem_validation&leads}b) resulted in a configuration which limited the field perturbation to 10.9~ppm. The optimized current leads are spread over three turns and arranged such that the inner and outer leads mutually have cancelling field contributions.

FEM simulations on COMSOL® were also carried out to validate analytical models used to calculate transient temperature distributions inside the magnet windings during operations. Based on scaled laboratory experiments, temperature-dependent properties were set considering 77~K temperature-imposed boundaries on the radial inboard and outboard surfaces. The multi-layered winding (70~µm~Cu/60~µm~PET, 150~turns) was modeled as solid parts with equivalent properties. The simulated time was set to 5 minutes. The resulting temperature distribution highlighted relatively high gradients located at the extremities of the magnet (highest current densities). At these locations, the computed maximum spatial temperature increase is 55~K, and the peak coil temperature is 132~K. %A first order thermo-electrical model with highly conservative assumptions used below for the mechanical design of the coil showed a maximum spatial temperature gradient of 95~K (at equilibrium) and peak coil temperature of 313~K (adiabatic temperature increase after 5 min without external cooling).
These results validated custom-made analytical thermo-electrical models, validating their conservative assumptions and confirming limited coil deformation during operation.

\subsection{Error sensitivity analysis}
\begin{figure}[ht]
\centering
\includegraphics[width=\columnwidth]{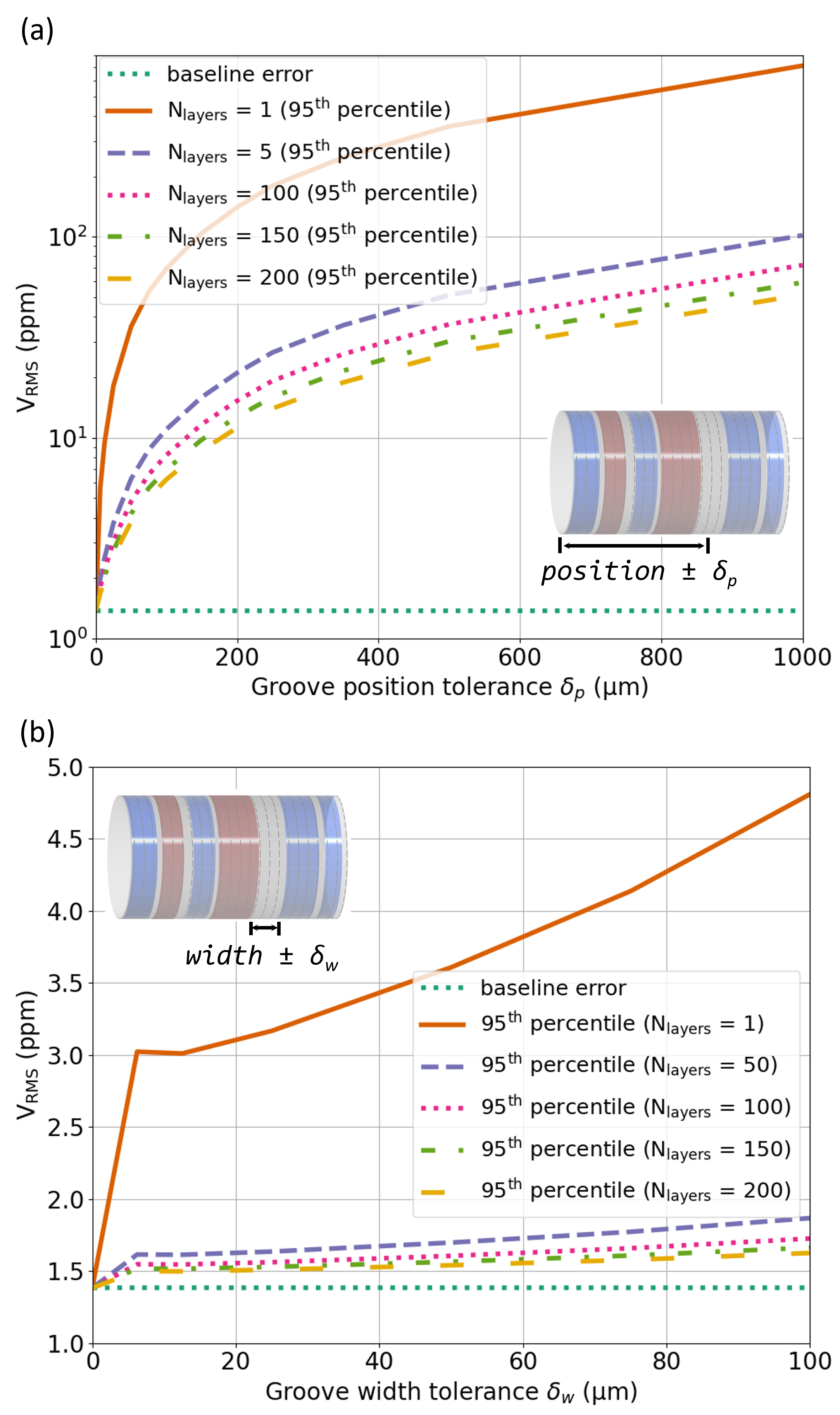}
\caption{Predicted effect on $V_\text{RMS}$ (vertical axis) of relaxing the construction 
tolerance (horizontal axis) on (a) the position and (b) the width of the grooves. 
Different curves refer to different numbers of layers, and are 
compared with the baseline $V_\text{RMS}$ (green dashed line). 
}
\label{fig:MRI_error_analysis}
\end{figure}

An analysis of the impact of mechanical inaccuracies on the precision of the magnet field was also carried out. More specifically, variations of groove width and position were randomly introduced in the model across all the coils over several runs while the differences between calculated and target fields were quantified (Fig.~\ref{fig:MRI_error_analysis}). Similar procedures were deployed to quantify the sensitivity to misalignment between windings and to overall stack thickness variations. These were omitted here for the sake of brevity.

Results shown in Fig.~\ref{fig:MRI_error_analysis} indicate that, in order to ensure a precision lower than 10~ppm, groove width and position inaccuracies must be within 100~µm. A similar conclusion was drawn regarding the misalignment between layers (\textit{i.e.}, layer misalignment tolerances of $\pm$ 100~µm, at most). Finally, even a systematic increase/decrease of a few µm per layer can as well degrade the field precision by few ppm. In this case, however, one could increase or decrease the number of layers in order to achieve the desired overall stack thickness and compensate any resulting field magnitude deviation by means of the intensity of the excitation current. 

\subsection{Experimental validation}
To demonstrate the ability of the proposed magnet design approach to accurately replicate a highly homogeneous field configuration, a prototype of the MRI test case and a test bench were manufactured. Since the automated system allowing the laser-engraving and the winding of a (super-)conductor foil over a cylindrical substrate was already functional at the time of this experiment, a laser-engraved copper laminated sheet was used to manufacture the prototype magnet.

\subsubsection{Prototype design and assembly}

\begin{figure}[ht]
\centering
\includegraphics[width=\columnwidth]{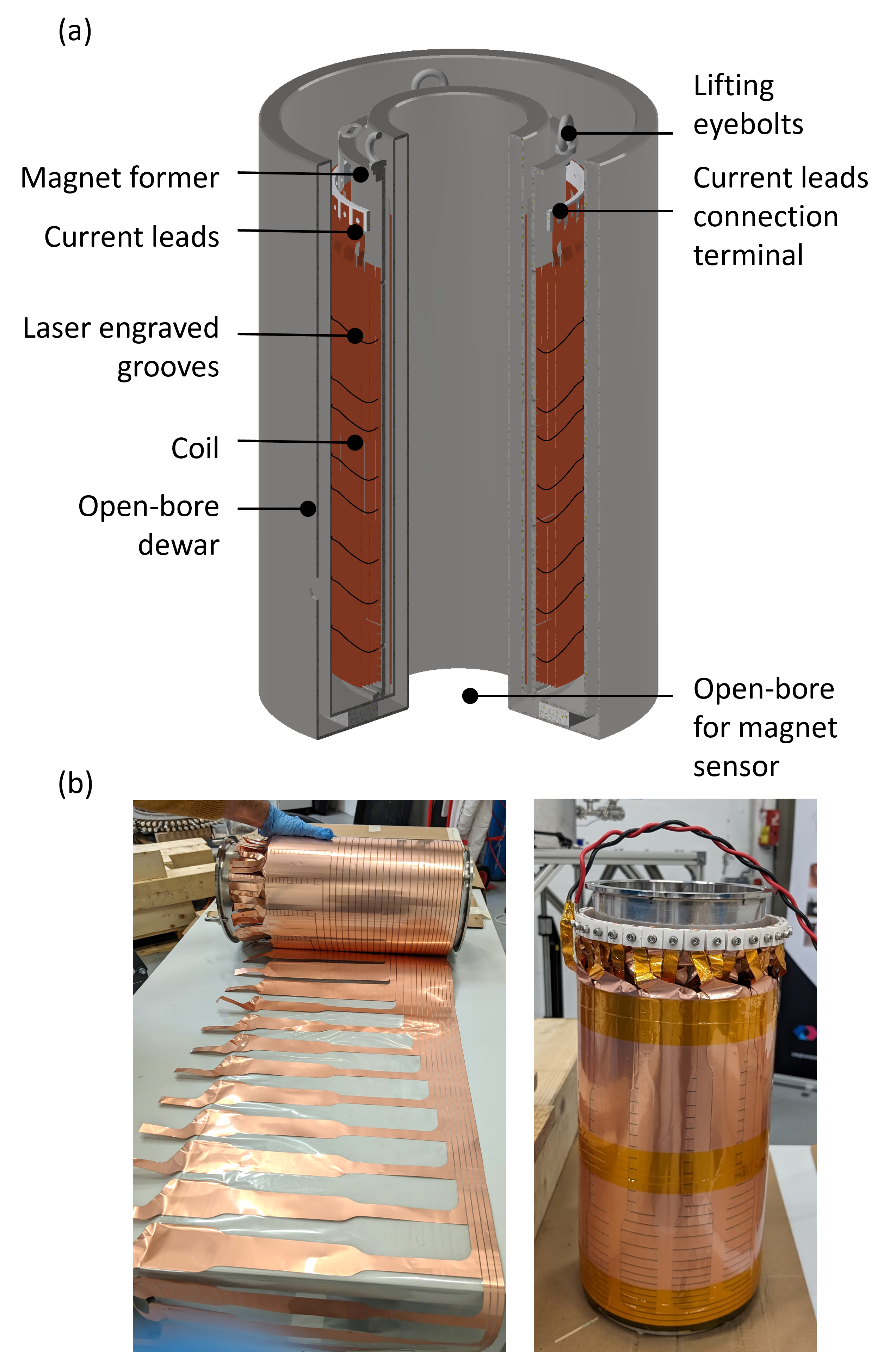}
\caption{(a) Cutaway view of the MRI magnet inside the open-bore dewar. (b) Prototype assembly showing the end turn with current leads (left) and the terminal connection ring (right).}
\label{fig:MRI_mec_prototype}
\end{figure}

The prototype mechanical design accounted for thermal calculations to ensure optimal performance and thermo-structural integrity of the MRI magnet system. The prototype magnet (Fig.~\ref{fig:MRI_mec_prototype}a) is made of wound copper laminated sheet of 70~µm~Cu and 60~µm~PET~+~adhesive which are laser-engraved with the design pattern (Fig.~\ref{fig:MRI_design}a) to produce the target uniform flux density of 0.1~T. The laser-engraving was tuned to cut through the copper layer without cutting through the PET layer to ensure mechanical integrity of the sheet during winding as well as the electrical insulation between layers. The entire magnet operated in liquid nitrogen inside a customized open-bore 316 stainless steel dewar (Fig.~\ref{fig:MRI_mec_prototype}a). Liquid nitrogen immersion was used to minimize Joule heating by leveraging the higher copper conductivity at such low temperatures. In order to reach the target field, minimize heating and material use, the prototype magnet was made out of 150 turns ($\sim$ 105 m of copper laminate). The laser engraved coils, powered in series, were connected through the innermost and outermost turn current leads at the top of the prototype through a terminal connection ring (Fig.~\ref{fig:MRI_mec_prototype}b). The designed magnet generates a peak power of 1.5 kW during operation requiring a refill of 0.5 liters per minute of liquid nitrogen.

\subsubsection{Measurement setup}
\begin{figure}[ht]
\centering
\includegraphics[width=\columnwidth]{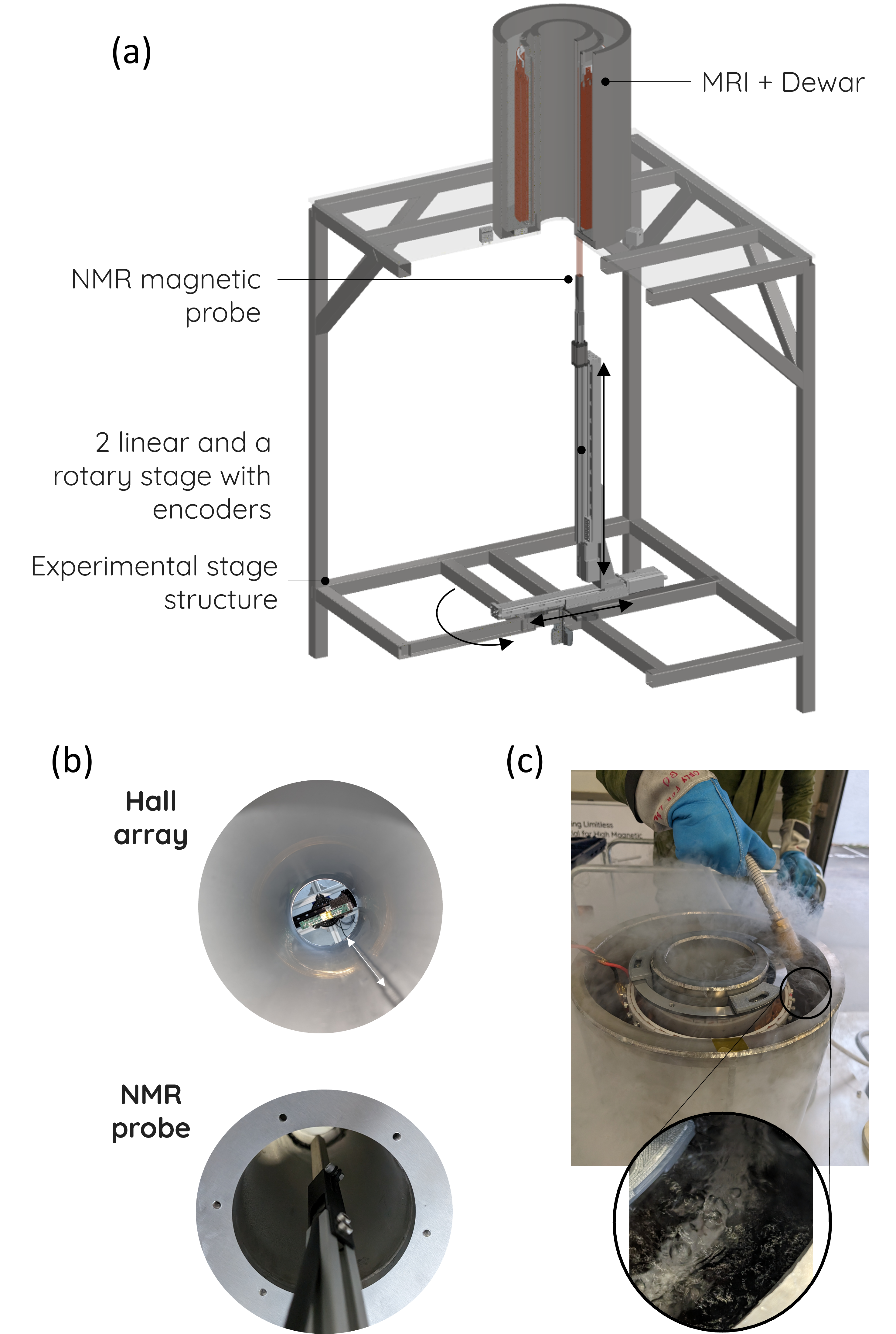}
\caption{(a) Experimental setup used for magnetic field mapping. (b) Photograph of the dewar internal bore fitted with (top) the hall sensor array, and (bottom) the NMR probe. (c) Photograph of the prototype cool down, immersed in a liquid nitrogen bath.}
\label{fig:MRI_measurement_setup}
\end{figure}

A PT2026 NMR sensor (MetroLab, Geneva, Switzerland), with specific range from 38 – 140 mT was mounted on a 3 degrees of freedom moving stage - two linear (Zaber) and a rotary plate (Thor Labs). The setup shown in Fig.~\ref{fig:MRI_measurement_setup} allowed for a 3D mapping of the magnetic field strength within the bore of the magnet in cylindrical coordinates. In addition, a hall probe array (Hallinsight from MetroLab) was also fitted in 
space of the NMR sensor to enable preliminary tests and ensure correct assembly. Most parts were made out of aluminium or 316 stainless steel to minimize field distortions due to ferromagnetic effects, and all 
ferromagnetic components such as the portable gantry crane, or the elevated platform used to work inside the dewar and refill liquid nitrogen were removed during measurements.

The magnet was powered by a 1 kV, 20 A (5 kW) power supply (MR100020, BK precision) through two flexible braided copper leads connecting to the terminal ring which, in turn, connects all the coils in series. The power supply provided a DC current up to a set point, thanks to a voltage upper limit. The applied current was measured through a 100~m$\Omega$ shunt resistor (1282 series, Burster) and a voltmeter (DMM7510, Keithley). Throughout the measurements, the device was powered for about five minutes at most to limit the temperature increase and boiling of the liquid nitrogen. The overall device temperature was monitored through the supplied voltage. 

The test campaign was organized into three phases. In the first phase, the magnet was cooled down by adding liquid nitrogen inside the dewar until it covered the prototype magnet and the terminal connection ring. Tracking the reduction in resistance allowed us to confirm the magnet temperature. In the second phase, the magnetic field along the axis and inside the VOI was measured using the hall probe array to ensure suitable alignment, operation and homogeneity before switching to the NMR sensor. During this phase, the temperature rise, sensor alignment and impact of the current leads were carefully monitored allowing for calibration and full setup of the device. In the third phase, the coils were powered at varying currents up to the nominal current and a full mapping of the magnetic field inside the VOI in the magnet bore was performed with the NMR probe. The moving stage displaced the NMR probe inside the VOI covering a circular area as shown in Fig.~\ref{fig:MRI_measurements}a. Measurements were then repeated for a $90^\circ$ rotated plane (azimuthal angle).

\subsubsection{Results}
\begin{figure}[ht]
\centering
\includegraphics[width=\columnwidth]{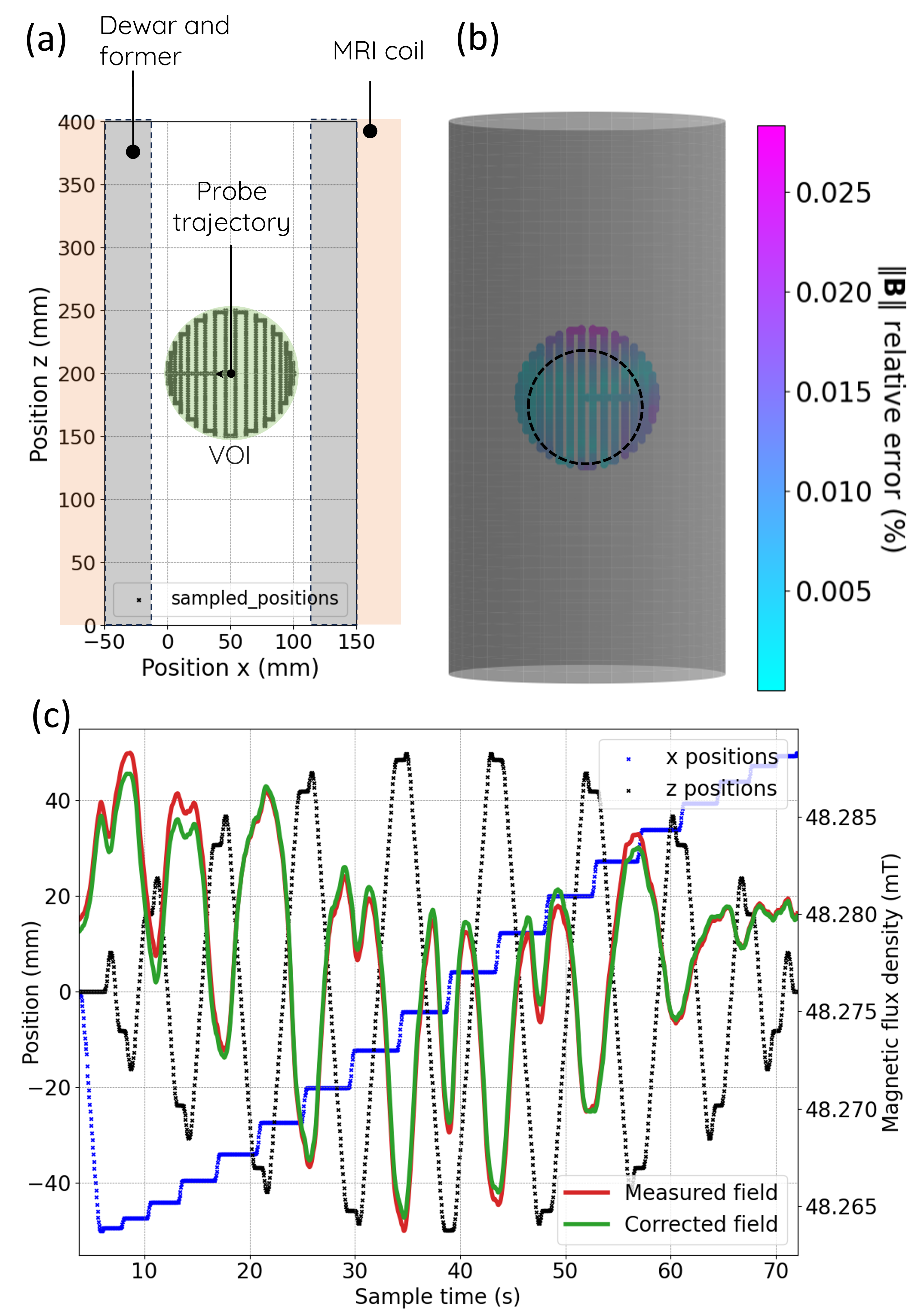}
\caption{(a) Illustration of the measurement setup and procedure highlighting the probe trajectory covering an initial 10 cm diameter VOI. (b) Measured relative error after post-processing: the dashed black circle highlights the area over which the higher field homogeneity was observed (8 cm diameter VOI). (c) Time-resolved measurements of magnetic field in the MRI magnet (right axis) as the NMR probe position is scanned (blue and black curves, left axis). 
Raw field measurements (red) were corrected (green) accounting for PS fluctuations over time. A basis current of 3.68 A - \textit{e.g.}, roughly half of the rate current for a 0.1 T field - was applied due to a better stability of the PS output at this range.}
\label{fig:MRI_measurements}
\end{figure}

The experimental data recorded from the NMR probe, linear stages and voltmeter are shown in Fig.~\ref{fig:MRI_measurements}. All the experimental data were resampled and realigned with respect to a reference time basis. Then, the voltage drop measured over the shunt resistor was used to correct field artifacts introduced by PS fluctuations over time (Fig.~\ref{fig:MRI_measurements}c).

Time-referenced experimental data (Fig.~\ref{fig:MRI_measurements}c) was then post-processed to show the spatial magnetic field map and the resulting local relative field error within the mapped area (Fig.~\ref{fig:MRI_measurements}b). The NMR probe was used to scan an area larger (10~cm DSV) than the volume of interest (8~cm DSV). A larger area was intentionally covered in order to allow for numerically aligning (along the axis and radially) the prototype magnet with the experimental setup (Fig.~\ref{fig:MRI_measurements}b). The measured local relative error was in the order of 10$^{-4}$. Table~\ref{MRI_results} summarizes the relative field error computation for three cases: theoretical model accounting for current leads interference, \enquote{as-built} model accounting for manufacturing tolerances and confirmed mechanical deviations. One can observe that measured error falls within the expected range, validating therefore not only models, analyses and tools developed to design the magnet but also the predicted sensitivity to experimental uncertainties. In addition, the experimental $V_{\text{RMS}}$ field homogeneity estimated over a 8 cm DSV also falls within the predicted range.

\begin{table}[h!]
\centering
\caption{Field homogeneity for a VOI with 8 cm of radius.}
\begin{tabular}{cccc}
\hline
 & \textbf{Model} & \textbf{As built} & \textbf{Measured} \\
\hline
$\varepsilon_{\text{max}}$ (ppm)& 40 & 150 $\pm$ 70 & 151 $\pm$ 10\\
$V_{\text{RMS}}$ (ppm)& 12 & 42 $\pm$ 12 & 52 $\pm$ 9\\
%$\varepsilon_{\text{max}}$ (ppm)& 40 & 80 - 220 & \textbf{152} \\
%$V_{\text{RMS}}$ (ppm)& 12.3 & 29 - 54 & \textbf{40.5} \\
\hline
\end{tabular}
\label{MRI_results}
\end{table}

\section{Conclusions and future work}

A new approach to simplify magnet design and manufacturing based on the patterning of large conductive or superconductive surfaces was introduced. Numerical methods were presented for optimizing grooved axisymmetric coils connected in series or parallel. Experimental proofs of concept validated the approach with scaled-down gyrotron and Magnetic Resonance Imaging (MRI) coil-sets connected in series.

While the gyrotron case demonstrated the approach’s ability to reproduce highly specific field configurations, the MRI prototype primarily highlighted its exceptional precision, achieving 10$^{-5}$ field accuracy. This level of accuracy exceeds the typical requirements for stellarator magnets \cite{otte_2016}, paving the way for designing these complex, non-axisymmetric magnets.

Extensive sensitivity analysis shed light on the sources of field errors and led to the development of error field compensation and minimization schemes. It is also essential to remark that a full-scale, 5$\times$ larger MRI magnet realized through the same manufacturing process would be 5$\times$ more precise. This is due to the mechanical tolerances remaining the same in absolute terms, but 5$\times$ smaller in relative terms. Furthermore, commercial MRI devices rely on \textit{a posteriori} shimming to achieve homogeneity of few ppm \cite{blasche_magnet_2017}, whereas our technology can potentially achieve this level without requiring such adjustments. This also suggests the potential for even better performance through \textit{a posteriori} error-field correction schemes.

Finally, the experimentally achieved homogeneity (52~$\pm$~9 ppm) was consistent with an \enquote{as built} numerical model (42~$\pm$~12~ ppm) incorporating realistic features such as non-axisymmetric current leads and observed mechanical inaccuracies. However, there is still room for design and construction improvements with respect to the numerical model results (12 ppm), which will be the subject of future work.

The proposed technique has the potential to significantly simplify the design of specialized magnetic devices and unlock opportunities for new applications. Beyond its relevance to gyrotrons and MRI magnets, wide superconducting surfaces could be patterned to create lossless busbars, energy storage devices, motors/generators, undulators \cite{bortot_2024}, and more.

The potential impact is expected to be particularly significant in the context of stellarator coil design. By transferring the complexity from the coil geometry to the current distribution, we foresee a major advancement in the design and fabrication of high-field stellarator magnets. Our approach is expected to result in a more streamlined manufacturing process, reduced costs, and improved performance of stellarator devices, which are crucial in advancing the pursuit of practical nuclear fusion energy.

\section*{Acknowledgements}
The authors would like to thank all Renaissance Fusion team and in particular some colleagues who directly participated in this project providing insightful advice or technical assistance: L.~Bortot, C.~Sborchia, Q.~Salquebre, G.~Scarantino and V.~Nicolas. Also, we would like to thank our collaborators from the StellaCage group, in particular M.~Sigalotti and Y.~Privat.

\bibliographystyle{IEEEtran}
\bibliography{bibliography.bib}
\end{document}